# A new way to detect the Higgs

**S Reucroft, Y Srivastava, J Swain and A Widom**

Department of Physics, Northeastern University, Boston, USA

E-mail: stephen.reucroft@cern.ch

**Abstract**. We describe a new technique to look for evidence of the Higgs mechanism. The usual method involves seeking evidence for the Higgs boson either directly or via the indirect effect that a virtual Higgs boson would have on a variety of Standard Model parameters. The new technique looks for Higgs field effects that are predicted to reduce the masses of heavy particles when they are in the presence of other heavy particles.

## 1. Introduction

In this article we describe a new way to look for evidence of the Higgs mechanism in experiments at very high energy. Details are given in two papers that have been posted on the LANL arXiv for several months. The first [1] describes the development of the theoretical ideas and the second [2] gives an overview of the experimental consequences.

The Higgs mechanism provides a mathematical technique for generating the masses of fundamental particles in the Standard Model. It has two major characteristics that can be used by experimentalists seeking evidence either for or against it. It predicts the existence of a Higgs boson with unknown mass and width and it predicts the existence of a ubiquitous Higgs field.

To date, all experimental methods seeking to confirm (or otherwise) the Higgs mechanism look for evidence of the **Higgs boson**, and these fall into two categories. The first seeks direct evidence of the Higgs boson via its production and decay characteristics. The second seeks indirect evidence via the effects of a virtual Higgs boson on a variety of Standard Model measurements.

The new technique described here depends on the existence of the **Higgs field** and the fact that heavy particles affect that field and thus the masses of other nearby heavy particles. This new method also has the advantage of being essentially independent of the Higgs boson mass.

## 2. Mass of the Higgs boson

In direct searches for the Higgs boson, nothing has ever been found and this allows a mass lower limit of 114.4 GeV/$c^2$ to be established. The LEP Electroweak Working Group [3] has performed complex fits to the available electroweak data assuming that a virtual Higgs boson contributes via loop diagrams. In these fits the assumed Higgs boson mass is an unknown parameter to be determined. Using this technique, the current preferred value for its mass is $85^{+39}_{-28}$ GeV/$c^2$, almost a standard deviation below the lower limit. Taking these two results together, one obtains a Higgs boson mass between 114 GeV/$c^2$ and 199 GeV/$c^2$ at the 95% confidence level. This does not prove that the Higgs boson exists, but it does give a strong indication that there is some phenomenon yet to be discovered. It might be the Higgs mechanism or it might be something else outside the Standard Model. This all makes it very likely that there is something exciting to be discovered at the LHC (turning on in 2007).

## 3. Higgs boson searches at the LHC

The two general purpose detectors at the LHC are Atlas and CMS. They have both made extensive simulation studies of potential Higgs signals in a variety of decay channels and with assumed Higgs boson mass up to 1 TeV/c$^2$. If there is a Higgs boson in the mass range between 114 and 199 GeV/c$^2$, as the indirect techniques indicate, then one of the best channels for discovery is Higgs $\rightarrow \gamma\gamma$. It will not be easy. For example, figure 1 shows a simulated $\gamma\gamma$ invariant mass plot for an assumed 130 GeV/c$^2$ Higgs boson using the CMS electromagnetic calorimeter [4]. The signal constitutes about 10 standard deviations and corresponds to an integrated luminosity of 100 fb$^{-1}$. With an instantaneous (and continuous!) luminosity of $10^{33}$ cm$^{-2}$s$^{-1}$, it would take ten years of stable operation to produce such a signal.

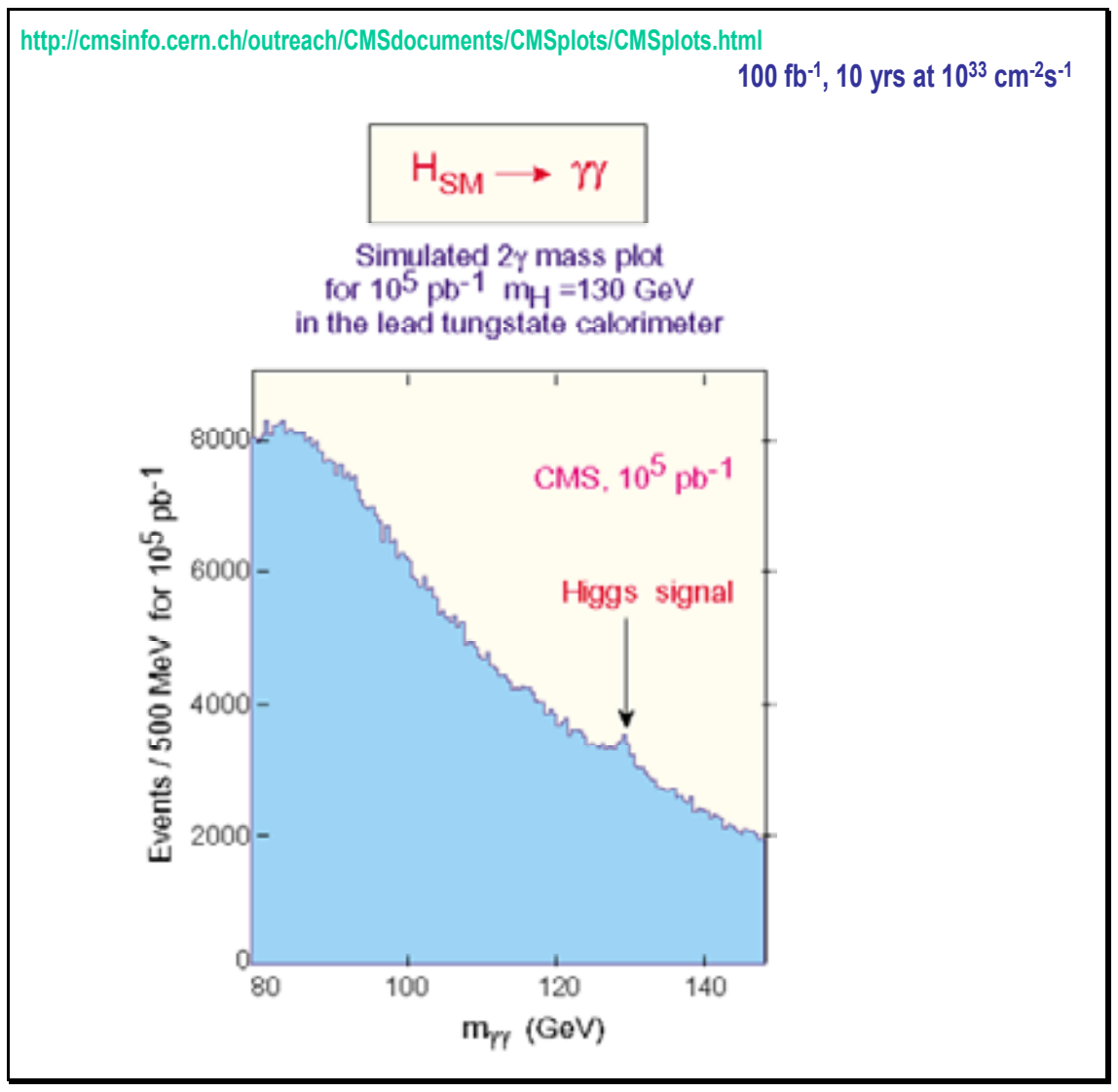

**Figure 1.** Simulated Higgs $\rightarrow \gamma\gamma$ mass plot for a 130 GeV/c$^2$ Higgs boson at CMS.

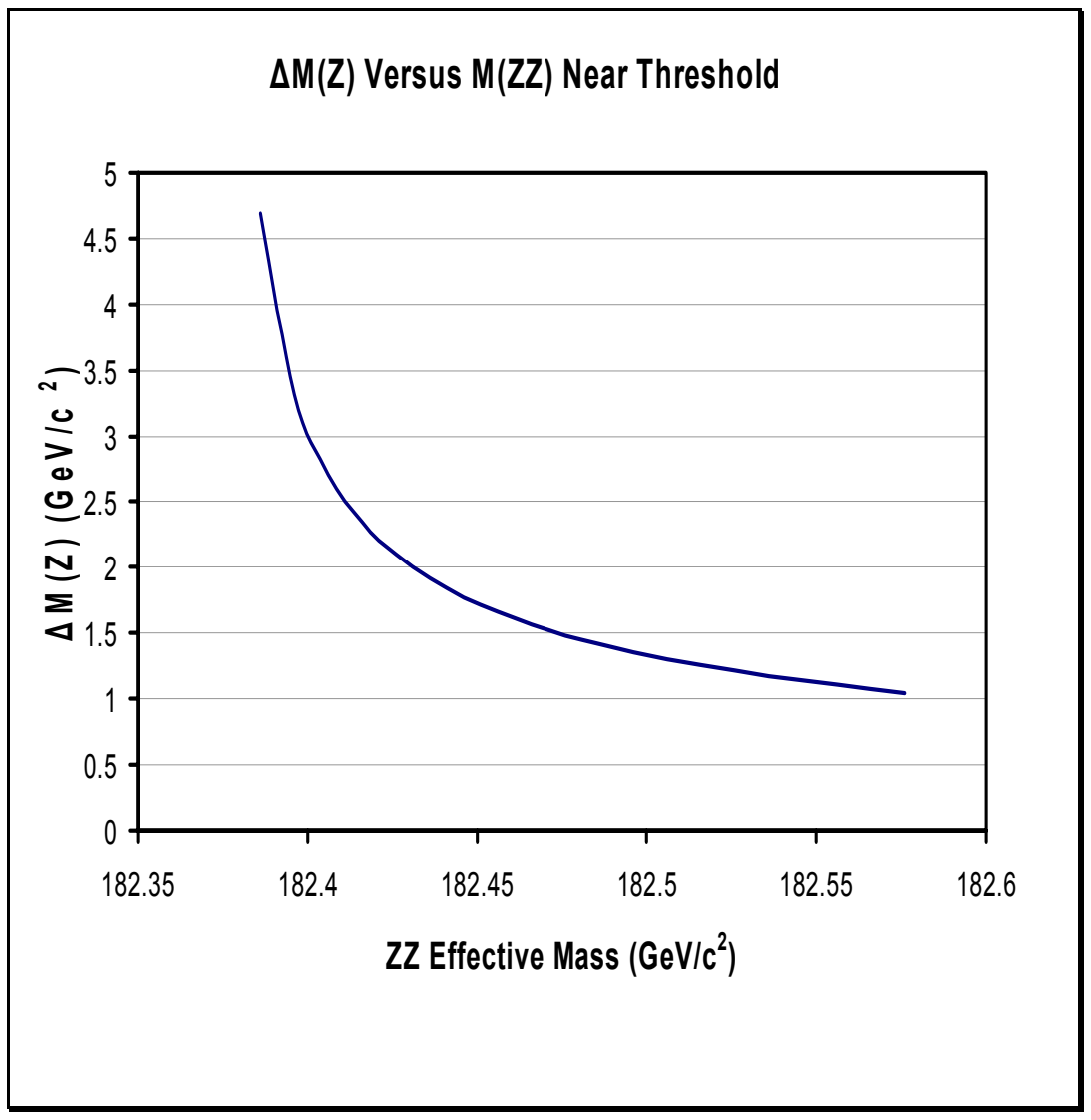

**Figure 2.** $Z$ mass reduction versus $ZZ$ invariant mass near threshold.

## 4. Massive particles as sources and detectors of the Higgs field

Motivated by the difficulty of direct searches for the Higgs boson, we have considered an alternative technique.

In two recent papers [1, 2] it was pointed out that high mass particles can be used as detectors of the Higgs field. This possibility arises because high mass particles constitute significant sources of the Higgs field and the resulting source-modified field can be detected by other high mass particles that suffer induced mass shifts.

The relevant formula provides the shift ($\Delta M_X$) in mass ($M_X$) of particle $X$ as a function of the invariant mass ($M_{XX}$) of the pair of particles, where $\Gamma_X$ is the width of particle $X$ and $v$ is the Higgs vacuum expectation value:

$$\Delta M_X \approx -\Gamma_X \left(\frac{M_X^2}{2\pi v^2}\right)\left(\frac{M_X^2}{M_{XX}}\right)\left(\frac{1}{M_{XX}^2 - 4M_X^2}\right)^{\frac{1}{2}} \ln\left(\frac{M_X}{\Gamma_X}\right).$$

Figure 2 shows the $Z$ mass reduction near $ZZ$ threshold. It is a significant effect. For example, within 100 MeV/c$^2$ of threshold the $Z$ mass is always lowered by more than 1.5 GeV/c$^2$; the $ZZ$ invariant mass has to be more than 17 GeV/$c^2$ above threshold before the $Z$ mass shift goes below 100 MeV/$c^2$.

Figures 3 and 4 show similar effects for the $W$ mass near $WW$ threshold and $t$ mass near $tt$ threshold. Again, the effects are significant.

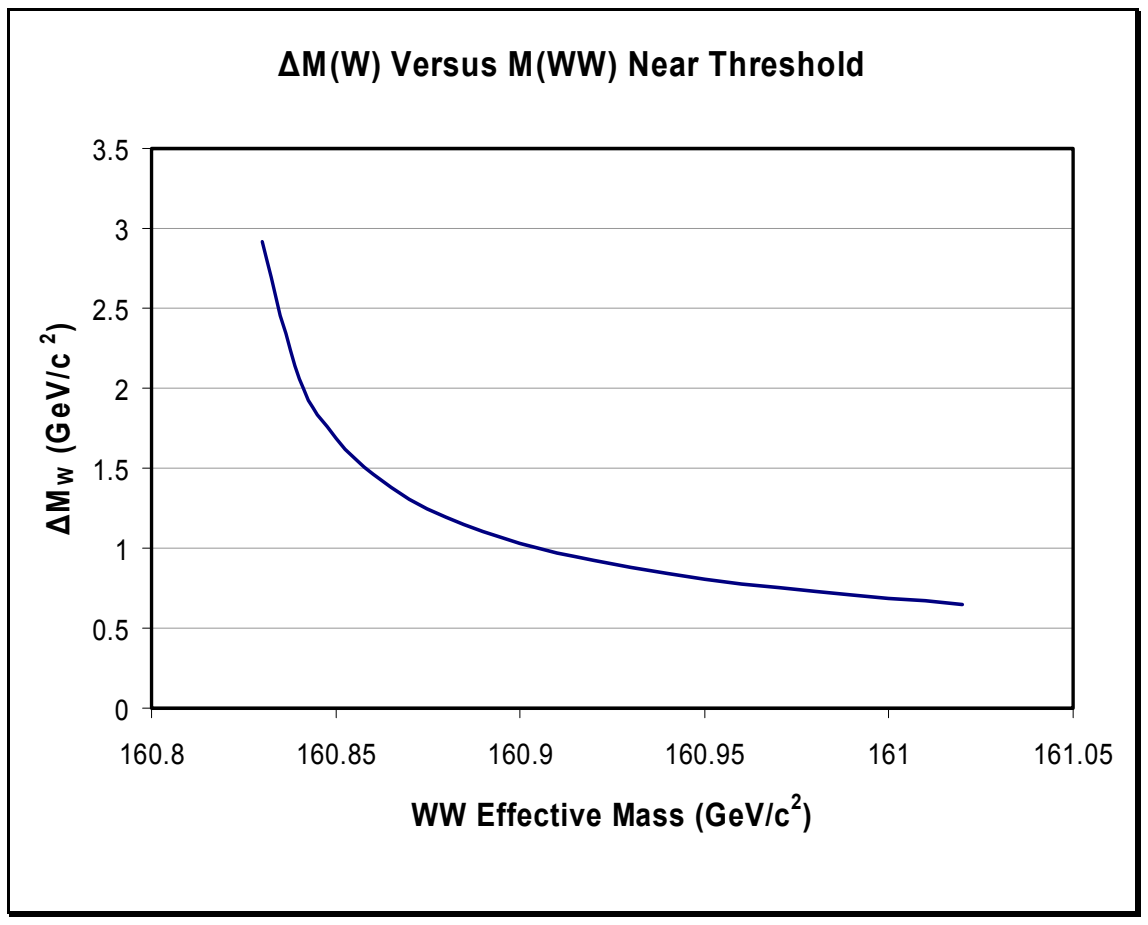

**Figure 3.** $W$ mass reduction versus $WW$ invariant mass near threshold.

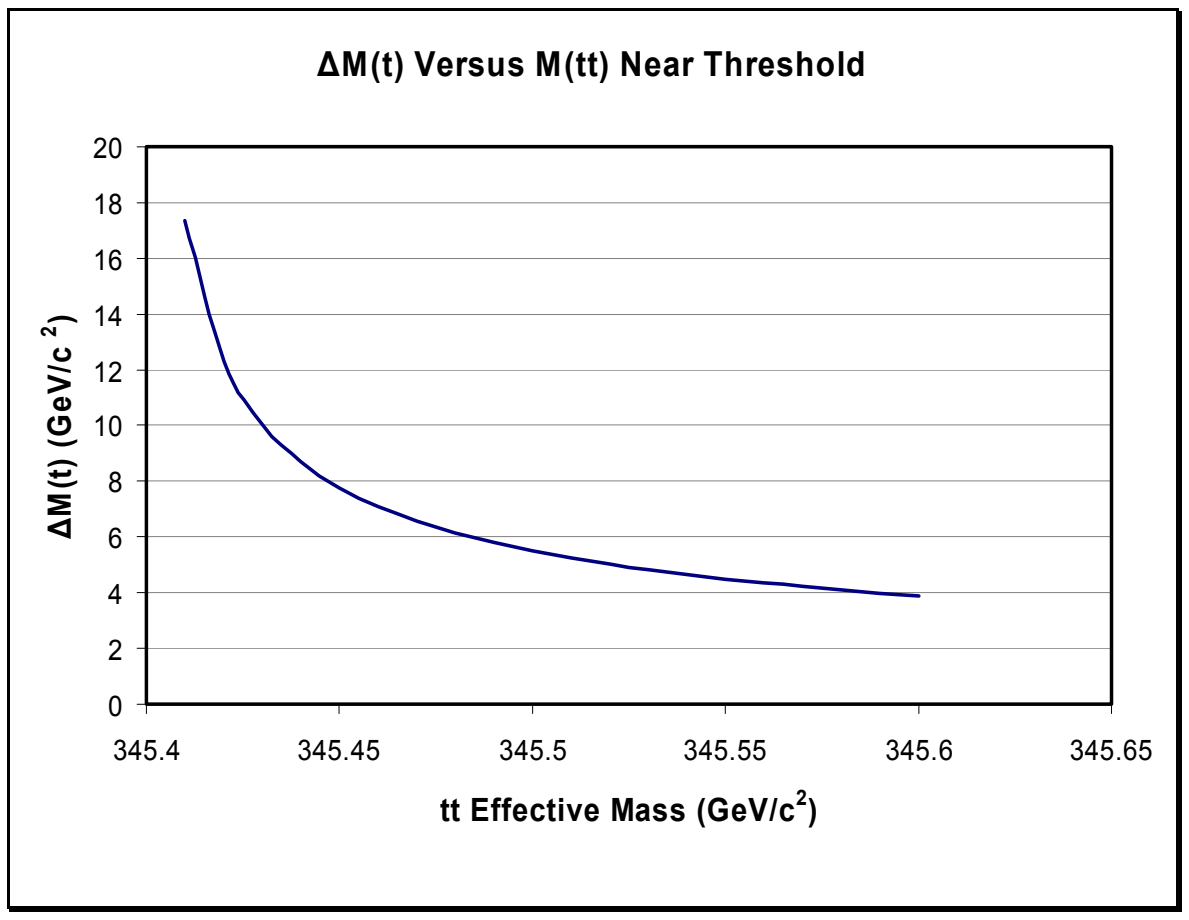

**Figure 4.** $t$ mass reduction versus $tt$ invariant mass near threshold.

## 5. Overview of *Z*, *W* and *t*-quark masses

The mass of the $Z$ has been determined very precisely at LEP1, the current world average value being $M_Z = 91.1876 \pm 0.0021$ GeV/$c^2$ [5]. At LEP2, once $ZZ$ threshold is reached, the $Z$ is produced in association with another $Z$ and the Higgs field related mass reduction [1] should be observable. Indeed, at the lowest LEP2 energy point above ZZ threshold, $E_{CM} \sim 183$ GeV, the predicted $Z$ mass reduction is greater than 800 MeV/$c^2$. Even at the next LEP2 energy point, $E_{CM} \sim 189$ GeV, the predicted $Z$ mass reduction is ~ 200 MeV/$c^2$. Unfortunately, not enough LEP2 data were collected at these near-threshold points to investigate the effect. As far as we know, none of the LEP experiments actually made an independent $M_Z$ determination using LEP2 data. The L3 experiment did publish $Z$ mass plots for different LEP2 energies [6, 7, 8, 9]. The low energy data clearly show truncated mass plots (presumably caused by the effect of threshold). Mass determination near threshold is not trivial! The most one can say is that it appears that $M_Z$ at LEP1 is consistent with $M_Z$ at LEP2; the predicted weighted average for the $M_Z$ reduction over the full LEP2 energy range would be ~ 140 MeV/$c^2$.

The mass of the $W$ has also been determined quite precisely at LEP2 ($WW$ pair production) and at the Tevatron (single $W$ production). The results are consistent and yield a world average of $M_W = 80.403 \pm 0.029$ GeV/$c^2$ [5]. The predicted average difference in $M_W$ from LEP2 and Tevatron data is on the order of the experimental uncertainty. Again, it would have been interesting to make a "precise" determination of $M_W$ at the lowest LEP2 energy point above $WW$ threshold where the predicted mass reduction is ~ 400 MeV/$c^2$.

Curiously, a careful determination of $\sin^2\theta_W$ by the NuTeV collaboration [10] combined with the world average value of $M_Z$ gives a value of $M_W$ some 3 standard deviations below the world average value of $M_W$. Since at NuTeV single $W$'s are produced by neutrinos, this is the opposite sign of the predicted Higgs field related mass shift.

The mass of the $t$ has been determined at the Tevatron, where the $t$'s are presumably produced in pairs, and the world average value is $M_t = 174.2 \pm 3.3$ GeV/$c^2$ [5]. There is no determination to date of $M_t$ in an environment where the $t$ is produced alone. It will be interesting, at future accelerator colliders, to carefully investigate events where the *t*-pair invariant mass is close to threshold. For example, at 10, 20, 30, 40 and 50 GeV/$c^2$ above threshold, the *t* mass is predicted to be shifted down by approximately 530, 360, 290, 240 and 200 MeV/$c^2$, respectively.

## 6. Summary

We describe the experimental implications of a new technique [1] for detecting the Higgs mechanism. Since it relies on Higgs field effects rather than Higgs boson effects it is essentially independent of the Higgs boson mass. The basic idea is to investigate the masses of heavy particles near pair threshold where the Higgs field of one particle reduces the mass of the other.

Of course it is a difficult and delicate technique, but it does provide a powerful Higgs mechanism tool for experiments at future very high energy collider accelerators, such as LHC, ILC, etc. and the technique does complement the more usual Higgs boson searches.

Unlike direct and indirect Higgs boson search techniques, it has the potential advantage, or perhaps disadvantage, of being able to provide unambiguous proof of the presence or absence of the Higgs field, whatever the mass of its associated Higgs boson.

## Acknowledgments

We thank our many colleagues on LEP, Tevatron and LHC experiments for many useful discussions and we are grateful to the NSF and the INFN for their continued and generous support of our activities.

## References

[1] Reucroft S, Srivastava Y, Swain J and Widom A 2005 *Preprint* hep-ph/0509151 (Accepted for publication in *Euro. Phys. J.*)

[2] Reucroft S, Srivastava Y, Swain J and Widom A 2005 *Preprint* hep-ph/0511233

[3] LEP Electroweak Working Group 2006 http://lepewwg.web.cern.ch/LEPEWWG/

[4] See http://cmsinfo.cern.ch/outreach/CMSdocuments/CMSplots/CMSplots.html for this plot and similar plots

[5] Particle Data Group 2006 *J. Phys.* **G33** 1-1232, http://pdg.lbl.gov and references cited therein

[6] Acciarri M *et al.* 1999 *Phys. Lett.* **B450** 281-93

[7] Acciarri M *et al.* 1999 *Phys. Lett.* **B465** 363-75

[8] Acciarri M *et al.* 2001 *Phys. Lett.* **B497** 23-38

[9] Achard P *et al.* 2003 *Phys. Lett.* **B572** 133-44

[10] Zeller G P *et al.* 2002 *Phys. Rev. Lett.* **88** 091802 and 2003 *Phys. Rev. Lett.* **90** 239902